# Lost chapter of Physical Chemistry means convergence between Fisher–Kolmogorov equation and tunnel effect.

Jiménez Edward [ab]*, Mosquera Héctor[a], Cortez Marco[a], Esteban Jiménez[b], Carlos E. Ayala[a], López Gustavo[a], Stahl Ullrich[a]

[a] Faculty of Chemical Engineering Central University of Ecuador, Quito, Ecuador.
[b] Microsoft Research, Quito, Ecuador. [c] Universite Paul Sabatier, Toulouse, France. * Corresponding author. Tel: +593994643752; E-mail: ejimenezecu@yahoo.com, ehjimenez@uce.edu.ec

Abstract

In this work we show that the dynamics of chemical reactions of order zero, one and two have a representation through logistics probability. This probability is robust, stable and complies systemically with the differential equation of Fisher-Kolmogorov (F-K).

It is robust, because in theorem 1 and theorem 3 differential equations of diffusion and heat transfer are obtained, where the temperature plays a key role. Also, the Eikonal equation of wave mechanics allows us to construct the heat equation. In Lemma 2, Fick´s diffusion equation is demonstrated.

It is stable, because probability convergence when t→∞, gives us new ways to analyze the kinetics of a reaction integrally, in Corollary 5.

Finally, the theoretically and experimentally obtained algorithms and results support the convergence in probability of the quantum tunnel effect and chemical reactions for: hydrogen production at ultra-low temperature and catalytic cracking of asphalt at high temperature.

**Keywords**

Fisher-Kolmogorov equation logistics/ quantum tunnel effect/, ultra-low temperatures,/ catalysis.

## Introduction

In Chemical Engineering, production or realization of the quantum tunneling effect necessarily involves the generation of reaction products, with a probability $P_+$ greater than zero, $P_+ > 0$. This probability must be in correspondence between the equations of quantum tunneling probability for which ($P_{tunnel}$) [1] is its solution and the probability $P_+$ of reaction kinetics theory. Firstly the solution of the Schrödinger equation $P_{tunnel}$ and the solution of the Fisher-Kolmogorov[1,2] equation $P_+$. Will be represented in the following way:

$$P_{tunnel} = \frac{1}{1 + e^{+2\pi|\epsilon|}} \qquad (1)$$

$$P_+ = \frac{1}{1 + e^{+kt}}. \qquad (2)$$

Where $\epsilon$ is an energy parameter[3,4], k represents the rate coefficient of Arrhenius equation and t is time.

These solutions show us the evolution of concentration of reaction products $C_+$ according to theorems (1), (3), (4) which will be demonstrated in next section. Furthermore, there exists a functional dependence between probability and products concentration[5-13]. The k parameter in solutions (1) and (2) are obtained from the zero-order ($C_+ = C_0 + kt$) or the first order ($\ln(C_+/C_o) = kt$) or the second order reactions ($1/C_+ - 1/C_0 = kt$). As an example, we can see a graphical representation of concentration and probability, in figure (1).

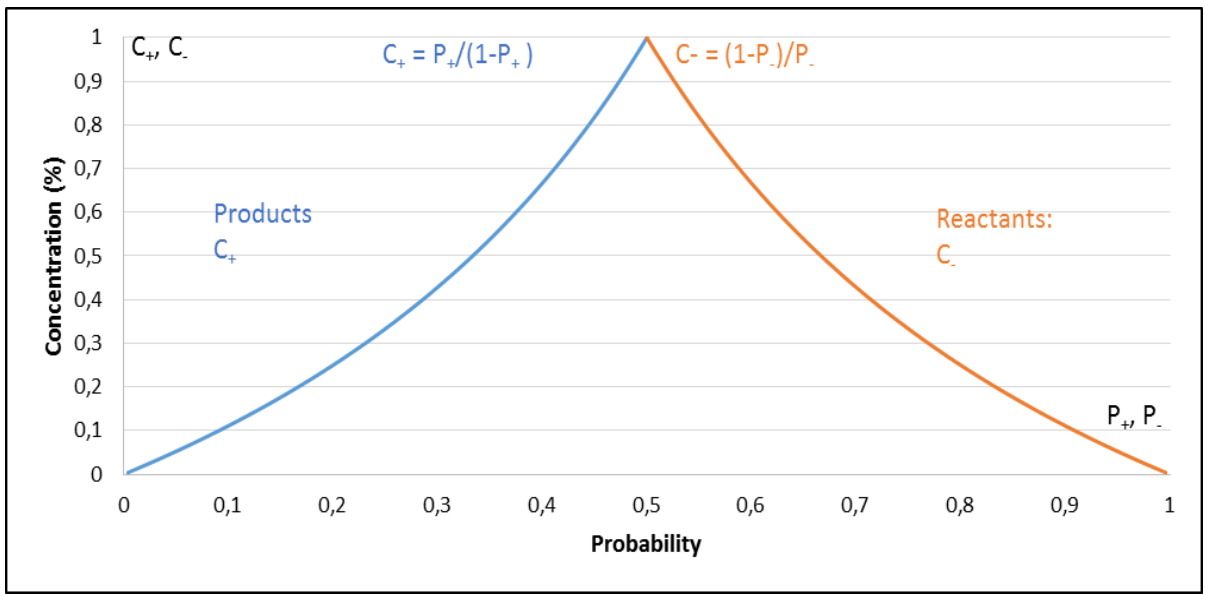

**Figure 1**. Showing the concentrations of reactants and products in function on the value of probability (blue: $C_+$ Vs $P_+$, orange: $C_-$ Vs $P_-$) for an irreversible first order reaction where reactants -> products.

**Chemical reactions and quantum tunnel effect have probabilistic behavior.**

The tunneling effect of chemical reactions occurs as well at temperatures below zero degrees Celsius as at temperatures above zero degrees Celsius. The nature of the tunneling effect is essentially quantum. In addition, in a chemical reaction the results are reaction products, which are macroscopically tangibles and measurable. Therefore, the mathematical formulation of the tunneling effect should have correspondence with classical obtained results. For the description of energy potentials, the Lennard Jones and Morse potentials[5] as inharmonic oscillators are used. For lower energy values of these potentials $E << V_{max}$ these show a good approximation to the harmonic oscillator model. The tunneling effect will be modeled with the harmonic potential which represents the energy potentials used in chemical reactions.

In a first example, we test Theorems 1, 3 and 4 in the hydrogen production by quantum tunnel effect, where we demonstrated with an average molar efficiency of 44.5%, at ultra-low temperatures using methanol in aqueous solution and oxygen. Both reagents interact in a batch reactor of a shell-tube type. As a coolant, liquid nitrogen was used. Visible light acts as photochemistry energy and both reaction classical probability $P_+$ and quantum tunnel probability $P_{tunnel}$ equations allows us to determine in a new manner the kinetic parameters of the reaction. The products of the reaction: $H_2$, $O_2$, and $N_2$, were determined via gas-chromatography.

For the production of hydrogen mostly steam reforming on light hydrocarbons is used; however this reaction requires high temperatures between 800 and 950°C, leading to high energy consumption, production of greenhouse gases as well as the use of non-renewable energy sources. Using catalysts it is possible to decrease the process temperature needed to levels around 350 °C.
Another approach is obtaining methanol through Fischer Tropsch synthesis,[14] which can be fed by renewable biomass energy sources.
Quantum Mechanics facilitates the modification of the known Arrhenius equation, allowing the incorporation of a tunnel effect in traditional chemical reactions. Thus, it is possible that a chemical reaction occurs without the necessary reaction conditions of the Classical Chemistry, in which the reactive compounds need to pass a barrier of energy-potential greater than the kinetic energy of the particles itself; this is due to low temperatures interactions at which this phenomenon is possible, as recently demonstrated,[15-19]. In addition, at the Princeton University, a study of the quantum tunneling was performed in chemical reactions, focused on NH3, [15-17].

**Quantum Arrhenius.**

Silva, V. et al. say: "Deviations from linearity in the Arrhenius plot lead to two very different regimes denoted by Sub-Arrhenius and Super-Arrhenius." [20]

The behavior of the regime Sub-Arrehenius in Figure 2 generates a concave curve by plotting (ln (k) vs 1 / T), this curve is characteristic of quantum phenomena.
In behavior regimen Super-Arrhenius in Figure 2 generates a convex curve which is characteristic phenomena where there are particles as diffusion transport and membrane permeability.

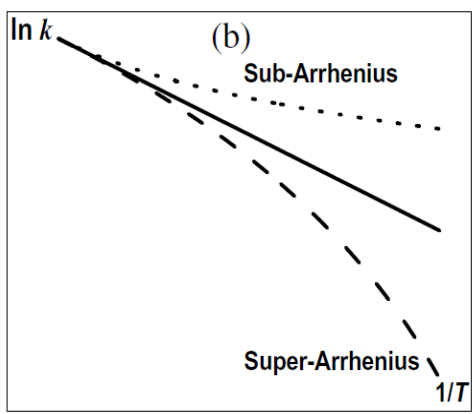

**Figure 2.** Linearity deviation Arrhenius[20].

By the other hand, R.P. Bell developed a quantum tunneling correction factor, Q, and explored its effect on an Arrhenius treatment of reaction kinetic, such as:

$$k_Q = QA_1 e^{-E_a/RT} \tag{3}$$

Where:

$$Q = \frac{e^\alpha(\beta e^{-\alpha} - \alpha e^{-\beta})}{\beta - \alpha} \tag{4}$$

With:

$$\alpha = \frac{E_a}{RT} \quad \text{and} \quad \beta = \frac{2a\pi^2(2mE_a)^{0,5}}{h} \tag{5}$$

This equation relates measurable reaction parameters to the probability of tunneling, allowing to experimentally determining if tunneling is taking place according to:

- Temperature Independence of the variable $k_Q$. We can see in Figure 3.
- Anomalous $E_a$ values.
- The independence of temperature leads to non-linear Arrhenius plots[21]

Calculations were performed by varying the temperature from 293 K to 54 K, in order to find the temperature value or range, where the tunnel effect is going to produce a reaction,In this case from 54 to 83 K, is the temperature range, the rate reaction constant of Arrhenius Quantum (k) becomes independent of temperature

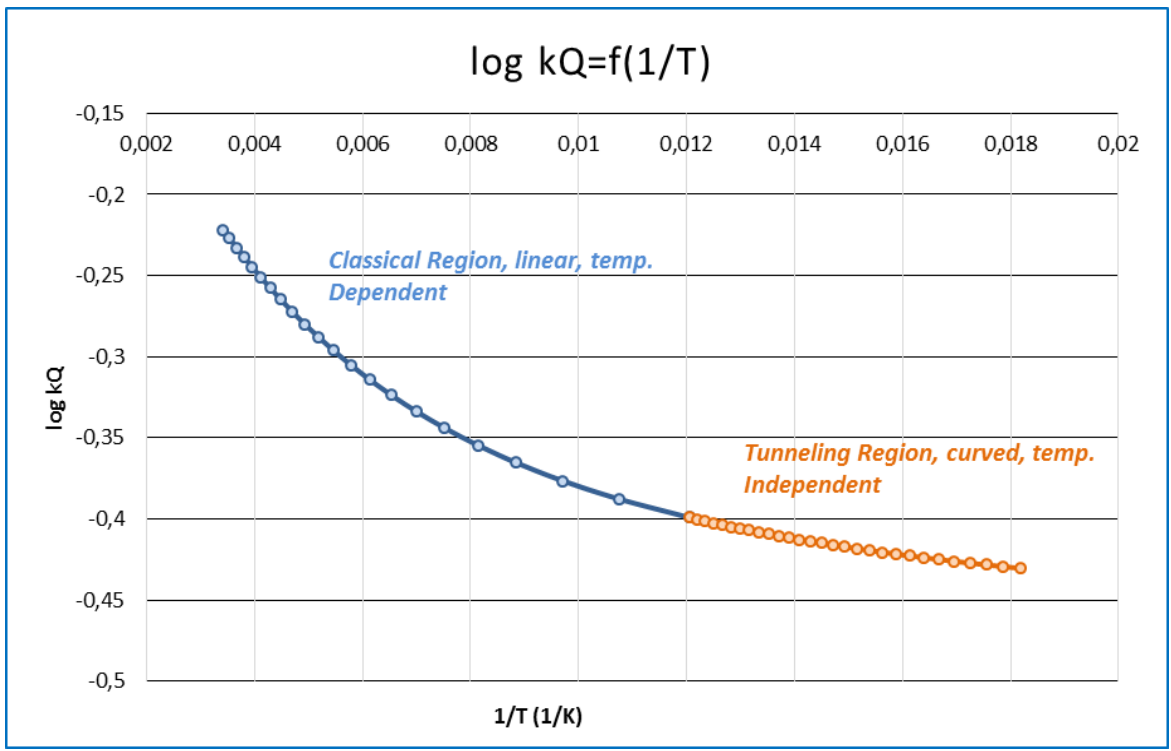

**Figure 3**. Arrhenius Quantum Equation. Theoretical and experimental foundations of hydrogen production. The temperature independence in the range (54, 83) K is showed (orange line); where the studied curve takes an evident linear trend. Furthermore, this theoretical set contains our experimental set of 82±1 K, giving us the coherence between theory and experiment.

The main subject of present work was to study photochemistry for the production of hydrogen from methanol based on the tunnel effect and at ultra-low temperature conditions, which is in the experimental interval of (82 ± 1) K, [19]. In this article, the theoretical and experimental bases of this phenomenon will be demonstrated, which could be a real achievement in the industrial production of hydrogen, due to the high energy demand and high economic costs required for the industrial production processes of hydrogen in the oil-producing countries and in the world [22-24]; it is necessary to implement a new method of obtaining hydrogen to achieve the reduction of the energy demand of oil and mitigate production of greenhouse gases, [25-27].

In a second example, we explore application of Fisher-Kolmogorov equation, Lemma 2 and Corollary 5 in the regeneration of nanoporous catalyst of refinery FCC unit, to crack and modify asphalt with the objective of a further recovery of lightweight composite.

Zeolitic catalyst regeneration was performed at 650°C, and it was characterized by BET surface area and infrared spectroscopy (FTIR). Catalytic cracking of asphalt and analysis of gaseous products were performed using thermo-gravimetric equipment and refinery gas chromatography. Principally, it was verified that the regenerated catalyst possesses special catalytic activity and tunnel effect, because a maximum percent of 11% recovery was obtained at a reaction temperature of 420°C and a very low weight ratio of catalyst / asphalt equal to 0,05; justifying the viability of tunnel effect when activation energy increases instead of decreasing.

The catalytic cracking[28-32] unit in fluidized bed (FCC) of Esmeraldas State Refinery (REE) is one of the most important in the current refining scheme due to its ability to convert heavy fractions of oil to products of higher economic value as gasoline. In this process large amounts of catalyst are used, which undergo deactivation due to coke deposition and other impurities on its surface, so they are removed and replaced [31-33]. This over spent catalyst is considered as a dangerous waste and represents a high environmental risk.

On the other hand, nowadays the demand for asphalt in Ecuador is very high due to the extension, repaving and construction of new roads, and it is a subject highly questioned for its quality and durability. It is also further known that the mechanical properties of asphaltic concrete are directly related to the physicochemical properties of asphalt, [33-36] so it is necessary to find alternative processes to improve their characteristics. In 2013 a group of researchers,[36,38] developed a thermocatalytic method to improve the Iranian empty residue using four different commercial catalysts, where they assessed the effect of temperature and relation catalyst/asphalt on the quality of both cracked products and the physicochemical properties of the residue.

There were three main objectives to develop this second example, the first is the regeneration and characterization of the spent catalyst of FCC unit, the second is to study the effect of temperature and the relationship catalyst/asphalt in a catalytic process and finally, the third, to analyze the occurrence of tunnel effect to high temperature.

The temperature at which the zeolitic catalyst was regenerated was 650° C [37,38] and it was characterized using analytical techniques such as infrared spectrometry by transformed of Fourier (FTIR), surface area and particle size distribution.

Zeolitic catalysts have higher efficiency at high temperatures, but it has to be taken into account, that the greater the temperature of catalytic cracking, higher the reactions of coke formation will be. [37,41]

So asphalt was subjected to a catalytic cracking by varying the temperature in a range from 400 °C to 440 °C and the weight ratio catalyst/asphalt in a range of 0,03 to 0,07; with the help of Thermogravimetry equipment (TGA). Finally the gaseous products were analyzed by Refinery Gas Analyzer (RGA).

THEORETICAL METHODS

**Definition:**

The Schrödinger equation, to be solved for harmonic potential has the form:

$$\left(\frac{d^2}{dx^2} + V^2(x)\right)\phi(x) = 0 \quad (6)$$

Where

$$V^2(x) = \frac{2m}{\hbar^2}\left(E - V_{max} + \frac{1}{2}\lambda^2\right) \quad (7)$$

Also, it is important to identify energy and frequency parameters used in solving differential Schrödinger equation, well developed in Nakamura [1]

$$\epsilon = \frac{E - V_{max}}{\hbar w^*} \quad (8)$$

$$w^* = \sqrt{\frac{\lambda}{m}} \quad (9)$$

The shape of the transmission probability from tunneling effect in the limit $E \ll E_a < V_{max}$ it is as follow.

$$P_{tunnel} = \frac{1}{1 + e^{2\pi|\epsilon|}} \tag{10}$$

Energy E in a chemical reaction may be thermal or electromagnetic.

**Definitions:**

The relationship between the rate coefficient k and temperature $T(x,y,z,t)$ is often represented by the Arrhenius equation[2] with $k = Ae^{-E_a/RT}$ where: the coefficient $E_a$, represents the activation energy. R the universal gas constant and T is the temperature in Kelvin degrees. The kt parameter corresponds to zero, one and second order reactions. And $k(T(x, y, z, t)$ has implicit dependence on space (x,y,z) and time (t). The Arrhenius equations and its concentration evolution $\partial C(P(t))/ \partial t$ are cornerstones of any chemical reaction.

**Theorem 1.** The probability of a chemical reaction P_( k(T(x,y,x,t)) complies Fisher Kolmogorov's equation and Eikonal[3] optics' equation, if and only if there is a function F( x,y,z,t) > 0 and $(\partial P\_) / \partial t < 0$.

$$\begin{cases} \frac{\partial P\_}{\partial t} - \nabla^2 P\_ = kP\_(1 - P\_); & 0.5 < P\_ < 0.823 \\ \frac{\partial T}{\partial t} - \nabla^2 T = 0; & (x,y,z) \in R; \ T, t \in R^+ \\ |\nabla T| = F(x,y,z,t) > 0; & (x,y,z) \in R; \ T, t \in R^+ \end{cases} \tag{11}$$

Where P_ is a logistic probability function, such that.

$$P\_ = \frac{1}{1 + e^{-kt}} \tag{12}$$

**Demonstration.**

We take the partial derivative with respect to time for the probability function P_(k,t), knowing that k = k (T(x, y, z, t) we can get:

$$\frac{\partial P\_}{\partial t} = tP\_(1 - P\_)\frac{E_a}{RT}k\frac{\partial T}{\partial t} \tag{13}$$

After, We take the Laplacian probability function P_(k(x ,y, z , t), t) to obtain:

$$\nabla^2 P\_ = t^2 kP\_(1 - P\_)(1 - 2P\_)|\nabla k|^2 + tP\_(1 - P\_)\nabla^2 \tag{14}$$

We can show that the laplacian probability depends on temperature. Laplacian and gradient of T(x, y, z, t), should be related to k (T) in the next manner.

$$|\nabla k|^2 = \left(\frac{E_a k}{RT}\right)^2 |\nabla T|^2 \qquad (15)$$

$$\nabla^2 k = \left(\frac{E_a}{RT}\right)^2 k|\nabla T|^2 + \frac{E_a k}{RT}\nabla^2 T - \frac{2E_a k}{RT^3}|\nabla T|^2 \qquad (16)$$

Replacing equations (15), (16) in (13) and (14) we have:

$$1 - \frac{E_a}{RT^3}\left[\frac{tE_a}{RT}\left[(1 - 2P_+)Ln\left(\frac{P_-}{1-P_-}\right) + 1\right] - 2\right]|\nabla T|^2 = \frac{tE_a}{RT}\left[\frac{\partial T}{\partial t} - \nabla^2 T\right] = 0. \qquad (17)$$

The equations (17) show us that diffusion equation and Eikonal equation occur naturally. Remember that Eikonal equation permit us to form heat equation.

$$\left[\frac{\partial T}{\partial t} - \nabla^2 T\right] = 0, \text{ and } 1 - F(x, y, x, t)|\nabla T|^2 = 0 \qquad (18)$$

Note that this equation of temperature distribution exists for P- ≠ 1; and k > 0.

The Eikonal equation exists only for the case of F(x,y,z, t) > 0, that implies values of 0.5 < P_ <0.82395909593032.

$$1 - \frac{E_a}{RT^3}\left[\frac{tE_a}{RT}\left[(1 - 2P_+)Ln\left(\frac{P_-}{1-P_-}\right) + 1\right] - 2\right]|\nabla T|^2 = 1 - F(x, y, z, t)|\nabla T|^2 = 0 \qquad (19)$$

The equation F(x, y, z, t) represents the probability limits and parameters specific to the chemical reaction.

Time variable denominate "t" may be replaced by the dimensions l and fluid velocities v within the reactor where the chemical reaction runs this is t = l / v.

Furthermore, all chemical reaction is defined in finite time (t << ∞), so time and temperature parameters are known and it is always possible to obtain F (x, y, z, t).

**Lemma 2.** The probability of a chemical reaction, $P_+$ ($k(T)$, t) meets the equation Fisher - Kolmogorov and the Eikonal equation if and only if there exists a function F ( x , y, z , t) > 0 with ( $\partial P_+$ ) / $\partial t$ > 0

$$\begin{cases} \dfrac{\partial P_+}{\partial t} - \nabla^2 P_+ = -kP_+(1-P_+); & 0 < P_+ < 0.5 \\ \dfrac{\partial T}{\partial t} - \nabla^2 T = 0; & x,y,z,T,t \in R^+ \\ |\nabla T| = F(x,y,z,t) > 0; & x,y,z,T,t \in R^+ \end{cases} \quad (20)$$

Where $P_+$ has a logistic probability function, such that.

$$P_+ = \frac{1}{1+e^{+kt}} \quad (21)$$

**Demonstration.**

We take temporally derivative and Laplacian from probability function, obtaining:

$$\frac{\partial P_+}{\partial t} = -tP_+(1-P_+)\frac{E_a}{RT}k\frac{\partial T}{\partial t} \quad (22)$$

In this equation changes the sign of each of the terms, if we compared to the time derivative $P_+$ that always has the decreasing over time.

$$\nabla^2 P_+ = -t^2 k P_+(1-P_+)(1-2P_+)|\nabla k|^2 - tP_+(1-P_+)\nabla^2 k \quad (23)$$

Taking the derivatives and gradients with respect to k.

$$|\nabla k|^2 = \left(\frac{E_a k}{RT^2}\right)^2 |\nabla T|^2 \quad (24)$$

$$\nabla^2 k = \left(\frac{E_a}{RT}\right)^2 k|\nabla T|^2 + \frac{E_a k}{RT^2}\nabla^2 T - \frac{2E_a k}{RT^3}|\nabla T|^2 \quad (25)$$

Substituting equation (24) and (25) in the fundamental equations (22) and (23), we obtain:

$$1 - \frac{E_a}{RT^3}\left[\frac{tE_a}{RT}\left[(1-2P_+)Ln\left(\frac{1-P_+}{P_+}\right)+1\right] - 2\right]|\nabla T|^2 = \frac{tE_a}{RT}\left[\frac{\partial T}{\partial t} - \nabla^2 T\right] = 0 \quad (26)$$

Equaling to zero each pair of equations that meet the Eikonal equations and diffusion in the range 0 < $P_+$ <1.

**Theorem 3.** Fick's law is a consequence of the probabilistic nature of a chemical reaction of order one, where. $C_+ = C_0 P_+ / (1 - P_+)$ and $C_- = C_0 (1 - P_-) / P_-$

**Demonstration.**

**Case 1.**

The concentration of one of the products of a chemical reaction has the form:

$$C_- = C_0 e^{-k(T)t} \tag{27}$$

In function of the probability we have:

$$C_- = C_0 \frac{1 - P_-}{P_-} \tag{28}$$

Taking partial derivative of $C_-$ with respect to time. In order to simplify calculations we can write $C_0 = 1$.

$$\frac{\partial C_-}{\partial t} = -\frac{1}{P_-^2} \frac{\partial P_-}{\partial t} \tag{29}$$

Now, if we find the laplacian of concentration.

$$\nabla^2 C_- = \frac{2}{P_-^3} |\nabla P_-|^2 - \frac{1}{P_-^2} \nabla^2 P_- \tag{30}$$

Equaling equations (29) and (30), we write.

$$\frac{\partial P_-}{\partial t} - \nabla^2 P_- - k P_- (1 - P_-) = -P_-^2 \left( \frac{\partial C_-}{\partial t} - \nabla^2 C_- \right) + \frac{2}{P_-} |\nabla P_-|^2 = 0 \tag{31}$$

Rearranging terms if $P_- \neq 0$, we have the first parenthesis that indicates Fick's second law.

$$\frac{\partial C_-}{\partial t} - \nabla^2 C_- = 0 \tag{32}$$

The resulting equation is the Eikonal from Classical Mechanics.

$$\frac{2}{P_-} |\nabla P_-|^2 + k P_- (1 - P_-) = 0 \tag{33}$$

There is a solution if and only if k < 0, which it is impossible. Knowing that k value is positive, the equation of the C_ concentration must be given a different formulation from the propose.

If we replace the concentration and concentration gradient:

$$|\nabla C_-|^2 + k\, C_-(1 + C_-) = 0 \tag{34}$$

From here we cannot find Fick's first law on positive real numbers.

**Case 2.** The concentration of one of the products of a chemical reaction has the form:

$$C_+ = C_0 e^{k(T)t} \tag{35}$$

In function of the probability we have:

$$C_+ = C_0 \frac{P_+}{1 - P_+} \tag{36}$$

Taking partial derivative of C with respect on time and to simplify calculations $C_0 = 1$

$$\frac{\partial C_+}{\partial t} = \frac{1}{(1 - P_+)^2} \frac{\partial P_+}{\partial t} \tag{37}$$

Now, if we find the Laplacian of concentration.

$$\nabla^2 C_+ = -\frac{2}{(1 - P_+)^3} |\nabla P_+|^2 + \frac{1}{(1 - P_+)^2} \nabla^2 P_+ \tag{38}$$

Substituting in the equation of probability.

$$\frac{\partial P_+}{\partial t} - \nabla^2 P_+ + k P_+(1 - P_+) = (1 - P_+)^2 \left( \frac{\partial C_+}{\partial t} - \nabla^2 C_+ \right) + \frac{2}{(1 - P_+)} |\nabla P_+|^2 = 0 \tag{39}$$

Rearranging terms and $P_- \neq 1$, we have the first parenthesis indicates Fick's second law.

(40)

$$\frac{\partial C_+}{\partial t} - \nabla^2 C_+ = 0$$

While the resulting equation is the Eikonal formulation.

$$\frac{2}{(1-P_+)}|\nabla P_+|^2 - kP_+(1-P_+) = 0 \qquad (41)$$

If we replace the concentration and concentration gradient:

$$|\nabla C|^2 - \frac{k}{2}\frac{C}{(1+C)^3} = 0 \qquad (42)$$

We can get Fick's first law as follows:

$$\sqrt{\frac{k}{2}\frac{C}{(1+C)^3}} = -\nabla C \qquad (43)$$

This equation has a solution if and only if k > 0.

In Hydrogen Production, theorem 3 is applied to obtain the activation energy depending on the elasticity (reaction rate constant versus temperature)

| N | CH$_3$OH/H$_2$O V/V | C$_{H2}$ % | P$_+$ | k$_i$t | k$_i$ | A$_i$ [1/s] | w$_i$ [1/s] |
|---|---|---|---|---|---|---|---|
| 1 | 99 | 29,7188 | 0,0326 | 3,3918 | 0,0565 | 98,8747 | 1,1373E+37 |
| 2 | 90 | 26,8113 | 0,0360 | 3,2888 | 0,0548 | 95,8734 | 1,1729E+37 |
| 3 | 80 | 26,4054 | 0,0365 | 3,2736 | 0,0546 | 95,4286 | 1,1784E+37 |
| 4 | 75 | 24,5337 | 0,0392 | 3,2000 | 0,0533 | 93,2854 | 1,2055E+37 |
| 5 | 70 | 27,8558 | 0,0347 | 3,3270 | 0,0555 | 96,9875 | 1,1595E+37 |
| 6 | 60 | 28,8213 | 0,0335 | 3,3611 | 0,0560 | 97,9807 | 1,1477E+37 |
| 7 | 50 | 25,5700 | 0,0376 | 3,2414 | 0,0540 | 94,4915 | 1,1901E+37 |
| 8 | 45 | 17,4881 | 0,0541 | 2,8615 | 0,0477 | 83,4170 | 1,3481E+37 |
| 9 | 40 | 45,8276 | 0,0214 | 3,8249 | 0,0637 | 111,5003 | 1,0085E+37 |
| 10 | 35 | 14,3932 | 0,0650 | 2,6668 | 0,0444 | 77,7392 | 1,4465E+37 |
| 11 | 30 | 25,2657 | 0,0381 | 3,2294 | 0,0538 | 94,1425 | 1,1945E+37 |
| 12 | 25 | 36,0742 | 0,0270 | 3,5856 | 0,0598 | 104,5241 | 1,0759E+37 |
| 13 | 20 | 42,1619 | 0,0232 | 3,7415 | 0,0624 | 109,0699 | 1,031E+37 |
| 14 | 15 | 19,6927 | 0,0483 | 2,9802 | 0,0497 | 86,8779 | 1,2944E+37 |
| 15 | 10 | 35,6978 | 0,0272 | 3,5751 | 0,0596 | 104,2184 | 1,079E+37 |
| 16 | 0 | 29,4986 | 0,0328 | 3,3843 | 0,0564 | 98,6579 | 1,1398E+37 |

**Table 1**. The first column indicates the volume ratio of methanol and water for the developed experiments. C$_{H2}$ % indicates the mole percentage of hydrogen produced in the chemical reaction of equation (72). Column 3, is an application of equation (36). The frequency $w_i$, is calculated using equation (9) of the tunnel effect.

We will write in a summarized way the algorithm on Table 1, so:

- Calculating the probability of the reaction depending on the concentration of hydrogen when, $C_0=1$.

$$P_+ = \frac{1}{1 + C_+} \quad (44)$$

Calculating the value $k_i t$ using equation (21).

$$k_i t = \ln\left(\frac{1 - P_+}{P_+}\right) \quad (45)$$

Knowing the time and temperature variables of our experiment we calculate the activation energy for t = 60[s] y T=82 K.

$$E_a = RT \frac{\partial \ln(k)}{\partial \ln(T)} = RT \frac{\frac{\Delta k}{k}}{\frac{\Delta T}{T}} = 5090 [\text{J/mol}] \quad (46)$$

Using data from Table 1, the standard deviation was obtained $\Delta k = 0,05$ for an average value of k=0,055; $\Delta T=1K$ and an operating temperature of T= 82 K, obtaining an activation energy $E_a$ =5090 J/mol, which is in similar work ranges (4200 -15000 )J/mol[19].

- Using equation (8) and equating equations (1) and (2), we can get the frequency $w_i$.

$$\epsilon = \frac{E - V_{max}}{hw^*} = k_i t \quad (47)$$

- the value of $k_i$ is obtained, for the time t=60 [s]
- the value of $A_i$ is obtained from Arrhenius equation. We have known values for T = 82 K, Ea = 5090 J / mol and R, it is the universal gas constant. We note that the value of Ai, is different for each concentration.

**Theorem 4.** Fick's laws are a consequence of the probabilistic nature of a chemical reaction of zero order and second order. $C_+ = C_0 + kt$ and $1/C_+ - 1/C_0 = kt$.

$$\frac{\partial C_+}{\partial t} - \nabla^2 C_+ = 0 \tag{48}$$

**Case 3.** Reaction zero order.

The concentration of one of the products of a chemical reaction of zero order has the form:

$$C_+ = C_0 + kt \tag{49}$$

Concentration depending on the probability is.

$$C_+ = C_0 + \ln\left(\frac{1-P_+}{P_+}\right) \tag{50}$$

Taking partial derivative of $C_+$ over time and to simplify calculations $C_0 = 1$

$$\frac{\partial C_+}{\partial t} = t\frac{\partial k}{\partial t} + k \tag{51}$$

Now, if we find the Laplacian of concentration.

$$\nabla^2 C_+ = t\nabla^2 k \tag{52}$$

Substituting gradient k and equaling equations (51) and (52).

$$k + t\frac{\partial k}{\partial t} = t\left(\left(\frac{E_a}{RT}\right)^2 k|\nabla T|^2 + \frac{E_a k}{RT^2}\nabla^2 T - \frac{2E_a k}{RT^3}|\nabla T|^2\right) \tag{53}$$

Deriving k versus time and replacing.

$$k + t\frac{E_a k}{RT^2}\frac{\partial T}{\partial t} = t\left(\left(\frac{E_a}{RT}\right)^2 k|\nabla T|^2 + \frac{E_a k}{RT^2}\nabla^2 T - \frac{2E_a k}{RT^3}|\nabla T|^2\right) \tag{54}$$

Rearranging terms, we have:

$$t\frac{E_a k}{RT^2}\left(\frac{\partial T}{\partial t} - \nabla^2 T\right) = -k + t\left(\left(\frac{E_a}{RT}\right)^2 k|\nabla T|^2 - \frac{2E_a k}{RT^3}|\nabla T|^2\right) \tag{55}$$

The first parenthesis indicates temperature diffusion equation.

And the second term is again the Eikonal equation for the temperature. Separating temperature's gradient, we obtain directly the heat equation, whose behavior is of a wave.

$$-k + t\left(\left(\frac{E_a}{RT}\right)^2 k|\nabla T|^2 - \frac{2E_a k}{RT^3}|\nabla T|^2\right) = 0 \tag{56}$$

**Case 4.** Reaction of second order

The concentration of one of the products of a chemical reaction of second order has the form:

$$\frac{1}{C} = \frac{1}{C_0} + kt \tag{57}$$

In function from the probability we have:

$$\frac{1}{C} = \frac{1}{C_0} + \ln\left(\frac{1-P}{P}\right) \tag{58}$$

Taking partial derivative of C + against to time.

$$\frac{\partial C}{\partial t} = kC^2\frac{E_a}{RT^2}\frac{\partial T}{\partial t} + kC^2 \tag{59}$$

Now, if we find the derivative of concentration and concentration Laplacian

$$\frac{\partial C}{\partial x} = -tC^2\frac{\partial k}{\partial x} \tag{60}$$

$$\nabla^2 C = 2t^2 C^3 |\nabla k|^2 - tC^2 \nabla^2 k \tag{61}$$

Substituting into the above equation the gradient an laplacian k equations (15) and (16), and building the Fick's equation we get.

$$ktC^2\frac{E_a}{RT^2}\frac{\partial T}{\partial t} + kC^2 = 2t^2 C^3\left(\frac{E_a k}{RT^2}\right)^2|\nabla T|^2 - tC^2\left(\left(\frac{E_a}{RT}\right)^2 k|\nabla T|^2 + \frac{E_a k}{RT^2}\nabla^2 T - \frac{2E_a k}{RT^3}|\nabla T|^2\right) \tag{62}$$

Rearranging terms, we have:

$$tC^2 \frac{E_a k}{RT^2}\left(\frac{\partial T}{\partial t} - \nabla^2 T\right) = -kC^2 - tC^2\left(\left(\frac{E_a}{RT}\right)^2 k|\nabla T|^2 - \frac{2E_a k}{RT^3}|\nabla T|^2\right) \quad (63)$$

The first parenthesis indicates the temperature diffusion equation.

$$tC^2 \frac{E_a k}{RT^2}\left(\frac{\partial T}{\partial t} - \nabla^2 T\right) = 0 \quad (64)$$

And the second term is the Eikonal equation for the temperature again.

$$kC^2 = 2t^2 C^3 \left(\frac{E_a k}{RT^2}\right)^2 |\nabla T|^2 - tC^2\left(\left(\frac{E_a}{RT}\right)^2 k|\nabla T|^2 - \frac{2E_a k}{RT^3}|\nabla T|^2\right) \quad (65)$$

Application to Catalytic Cracking and Tunnel Effect.- We analyze the applied models to build Table 2.

| Asphalt | Catalyst/Asphalt % | Catalysis Activity | Thermal Cracking | Activity Energy: Ea J/K.mol | Collission Frequency: A Hz | Tunnel Effect | $R^2$ Score | Error % |
|---|---|---|---|---|---|---|---|---|
| Sample 1 | 0,01 | No | Yes | …… | …… | …… | …… | …… |
| Sample 2 | 0,03 | No | Yes | …… | …… | …… | …… | …… |
| Sample 3 | 0 | No | Yes | 25518 | 1,2064435 | Not | 0,937 | 0,2438 |
| Sample 4 | 0,05 | No | Yes | 26272 | 1,4076609 | Yes | 0,963 | 0,1512 |
| Sample 5 | 0,05 | Yes | Yes | 25087 | 1,1748137 | Not | 0,962 | 0,1571 |
| Sample 6 | 0,05 | Yes | Yes | 24100 | 0,9730588 | Not | 0,966 | 0,1376 |
| Sample 7 | 0,07 | No | Yes | …… | …… | …… | …… | ….. |
| Sample 8 | 0,1 | No | Yes | …… | …… | …… | …… | ….. |
| Sample 9 | 0,15 | No | Yes | …… | …… | …… | …… | ….. |

**Table 2.** Analysis of the results for each sample analyzed. The activation energies of each sample and the setting value $R^2$ are calculated. For catalyst concentrations: 0; 0,01; 0,03; 0,07; 0,1; 0,15 There is no catalytic activity.

Sample 5, The catalyst has no catalytic activity because the activation energy increases with the activation energy of the thermal cracking; 26272>25518. However, higher yields are obtained in Cracking, which can be explained by tunneling.

the calculation of the concentration residue cracking is performed using the equations (12) and (28). This concentration explicitly dependent on time t and temperature T.

$$C_i = \frac{1 - \frac{1}{1+e^{-kt}}}{\frac{1}{1+e^{-kt}}} \tag{66}$$

The root mean square mistake is minimized and we found the value of the activation energy and the constant A, which are in column 5 and 6 of the table2.

$$Error = \sum_{i=1}^{n}(C_i - C_i^*)^2 \tag{67}$$

Where $C_i^*$ is the experimental value of concentration, R2 is obtained from the regression $(C_i, C_i^*)$ for i=1,66. According to the respective tables contained in the supplementary material.

**Corollary 5.** Convergence and stability in time from laws of: Fick, temperature diffusion and heat[42].

$$\lim_{t\to\infty}\frac{E_a}{RT}t\left[\frac{\partial T}{\partial t} - \nabla^2 T\right] = \left(\lim_{t\to\infty}\frac{E_a}{RT}\right)*0 = 0 \tag{68}$$

The diffusion equation is satisfied for t > 0.

The equation for the probability $P_+$ is equal to zero when t →∞. Due to which we have:

$$\lim_{t\to\infty}(kP_+(1-P_+))\left(1 - \frac{E_a}{RT^3}\left[\frac{tE_a}{RT}\left[(1-2P_+)Ln\left(\frac{1-P_+}{P_+}\right)+1\right]-2\right]|\nabla T|^2\right) = 0 \tag{69}$$

If the concentration C approaches zero when time t → ∞, then we have the trivial case.

$$\lim_{t \to \infty} kC^2 - 2t^2 C^3 \left(\frac{E_a k}{RT^2}\right)^2 |\nabla T|^2 = \lim_{t \to \infty} -tC^2 \left(\left(\frac{E_a}{RT}\right)^2 k|\nabla T|^2 - \frac{2E_a k}{RT^3}|\nabla T|^2\right) = 0 \quad (70)$$

If the concentration C <∞ and t → ∞, then:

$$\lim_{t \to \infty} kC^2 - 2t^2 C^3 \left(\frac{E_a k}{RT^2}\right)^2 |\nabla T|^2 = \lim_{t \to \infty} -tC^2 \left(\left(\frac{E_a}{RT}\right)^2 k|\nabla T|^2 - \frac{2E_a k}{RT^3}|\nabla T|^2\right) = -\infty \quad (71)$$

This result, demonstrate us compliance equality and the respective laws diffusion, Eikonal and heat transfer.

EXPERIMENTAL METHODS

**FIRST EXAMPLE(Tunneling effect at ultra-low temperatures)**

This part contains the experimental method for producing hydrogen by tunneling with an efficiency of 44.53% molar, having as reagents in methanol aqueous solution and oxygen. Both reagents interact in a transparent tube-shell type batch reactor, using liquid nitrogen as a refrigerant. Irradiating visible light is used to provide the activation energy of rupture of the C-H bond in the molecule of methanol and ultra-low temperatures as parameters required so that the probability of Tunneling is given and allow the photochemical reaction occurs. Quantum the Arrhenius equation to determine the kinetic parameters of the reaction. Products: $H_2$, $O_2$ and $N_2$ were determined by gas chromatography.

The experimental development conducted to obtain hydrogen, included a photochemistry reaction with a visible light source, using an aqueous methanol solution as reactive source using a partial oxidation process with oxygen (oxidizing agent) as detailed in the following empirical chemical reaction:

$$CH_3OH + H_2O + O_2 \rightarrow CO_2 + 3H_2 + O_2 \quad (72)$$

Although the empirical equation could be written as follows,

$$CH_3OH + \frac{1}{2}O_2 \rightarrow CO_2 + 2H_2 \tag{73}$$

The experimental process is summarized in Figure 4.

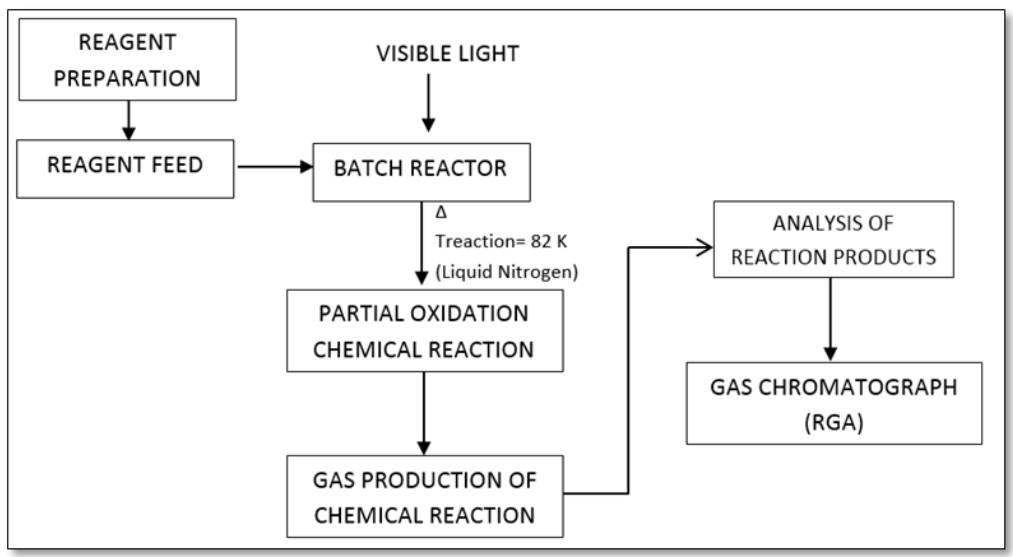

**Figure 4.** Flow diagram of the experimental setup.

**Batch reactor.** It is a reactor transparent with a design type tube-shell, on the tube side the reactants were fed in countercurrent and by the shell side liquid nitrogen as a refrigerant was fed to achieve the temperature at which the tunnel effect occurs ($82 \pm 4$ K). Using visible light LED bulbs, the photochemistry energy was provided.

**Reagent feed.** The aqueous methanol solution is injected in an atomized way by one end of the reactor, simultaneously at the other end gaseous oxygen is injected. Once you have finished feeding the required amounts of reagents a certain time is expected so that the reaction product into the reactor stabilize and can do the sampling.

**Sampling of the reaction product**. The tight syringe attached to the capillary was maintained into the reactor. An amount of sample approximately 20 ml is taken and the syringe valve closes to prevent leakage of the gaseous product; it´s immediately carried to injected into the loop feeding chromatograph.

**Analysis of the sample of the reaction product.** Into the gas chromatograph the analysis sample obtained from the chemical reaction is injected, it is done with a constant flow so that when the analysis is running on the chromatograph, the loop supply is full sample. At the end, the experimental spectra were obtained for: hydrogen, oxygen and nitrogen.

**Experimental Sample Analysis.**

- One, the chromatograph was fed with an air sample using a tight syringe to discard waste hydrogen into the syringe

- Two, the obtained sample was loaded from the chemical reaction, experimentally obtaining hydrogen, oxygen and nitrogen.

- Three, theoretical chemical reaction has as products hydrogen and carbon dioxide, however at the time of analysis of the sample the chromatograph fails to identify the signal of carbon dioxide since the operating temperature is too low to keep it in its gaseous state, because at that temperature $CO_2$ it condenses and subsequently solidifies on the walls of the reactor.

- Finally, products in the output of the reactor are not all part of the fundamental reaction, as follows:

➤ The excess oxygen is which did not react and of course it recovers in the output.

➤ Nitrogen appears because compressed air is mixed with the methanol reagent to be sprayed.

The two runs of both nitrogen and air were performed with the aim that at the moment to feed the sample, product of reaction, chromatograph's analysis lines are completely clean and can be measured with accuracy, unmistakably the existence of hydrogen in our sample.

# SECOND EXAMPLE (Tunnel effect to high temperature.)

This part contains the experimental method, which consists of two stages, the first is the regeneration and characterization of FCC catalyst and the second corresponds to the catalytic cracking of asphalt and analysis of gaseous products. As shown in the figure (3)

In the second stage, two independent variables were determined: the temperature and ratio mass catalyst / mass asphalt, and a dependent variable: the percent recovery of lighter compounds.

Were established the different ranges of operation for each of stages, through preliminary tests.

| Process | Variable | Range |
|---|---|---|
| Regeneration | Temperature (°C) | 650 |
| | Reaction time (min) | 30 |
| Catalytic Cracking | Temperature (°C) | 400–420–440 |
| | Ratio Catalyst/Asphalt $[g_{cat}/g_{asf}]$ | 0,03-0,05-0,07 |
| | Reaction time (min) | 30 |

**Table 3.** Variables and conditions of the experimental method

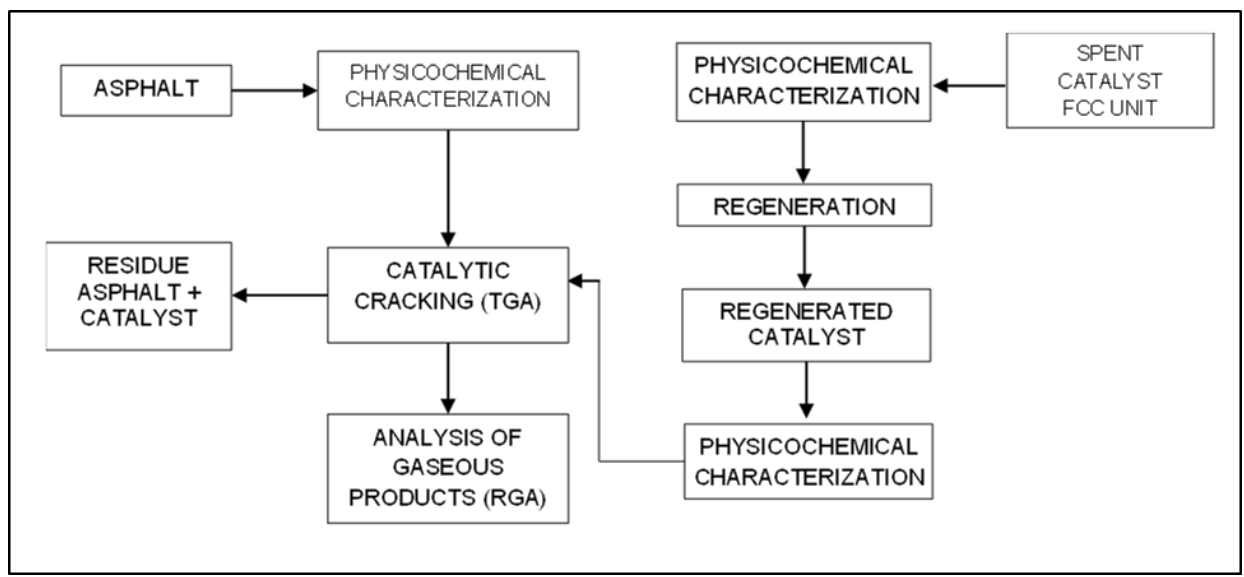

**Figure 5.** Flow diagram of the experimental setup to the catalytic process

**Catalyst:** A spent catalyst from the Fluid Catalytic Cracking Unit of Esmeraldas State Refinery in Ecuador was used.

**Asphalt:** The sample comes from the vacuum distillation unit State Esmeraldas Refinery.

a) Regenerate the FCC catalyst at 650 ° C.

b) Characterize the spent and regenerated catalyst by a surface area tests, infrared spectrometry Fourier transform (FTIR) and particle size distribution.

c) Determine the existent percentage of coke in the spent catalyst.

d) Take a sample of asphalt and determine the necessary theoretical amount of catalyst.

e) Perform the catalyst mixture by adding the asphalt, then reheated to a temperature between 60-80 ° C and stirring continued for a time of 10 minutes.

f) Create a method for each reaction temperature set in the Thermogravimetry equipment, and program a dynamic thermal process as shown in the following table.

| Conditions | Unit | Value |
|---|---|---|
| 25°C to T Reaction | °C/min | 10 |
| T ° Reaction | minutes | 30 |
| Inert atmosphere ($N_2$) | mL/min | 75 |

**Table 4.** Thermal Process catalytic cracking

g) Take a sample of the mixture and put it into the cell team for analysis.

h) Repeat this process for all asphalt relations *m/m* catalyst/asphalt and different reaction temperatures.

i) Identifying the catalytic cracking zone and calculate the area under the curve.

j) Calculate the percent recovery of lightweight composite asphalt.

k) Determine the optimum process conditions.

l) Analyzing the gaseous products for gas chromatography refinery.

# RESULTS AND DISCUSSION

The connection of the quantum world and the microscopic world occurs by stochastic processes and measurable parameters of the laws of probability. In exceptional cases the laws of probability are the same for classical and quantum effects.

Generally in a chemical reaction, the probability values which delivers the tunnel are similar to the probability values obtained by classical methods reaction-diffusion, which are well explained by the differential equation Fisher-Kolmogorov.

However, in our experiments the laws of probability and probability value are equal because we work at low energies, where the harmonic molecular potential is a reasonable approximation.

**1. Meaning and application of Theorem 1**.

Any chemical reaction has a classic component represented by the Fisher-Kolmogorov equation and a component quantum represented by tunneling.

Quantum reality, probabilistic essentially becomes visible and measurable in chemical reactions through the tunnel effect.

Theorem 1, is a systemic explanation of the logistics probability, therefore it allows simultaneously fulfill the fundamental equations of chemical reactions, that is, diffusion, heat and Ficks.

**2. Meaning and application of Lemma 2.**

The Lemma 2, It is a generalization of Theorem 1 applies to reagents, where concentrations increase C + and have a functional dependence with the probability P +.

**3. Meaning and application of Theorem 3 and of Theorem 4**

Fick's law is widely used to understand the dynamics of the concentration of products and reactants of a chemical reaction.

The diffusion equation, also called 'Fick's second law of diffusion', relates the rate of change of concentration at a point to the spatial variation of the concentration at that point.

The order of reaction determines the effective concentration of the probability of reaction as follows:

For zero-order reaction.

$$C = C_0 + kt;\ kt = \ln(\frac{1-P_+}{P_+}) \quad (74)$$

For first order reaction.

$$C = C_0 e^{kt};\ kt = \ln(\frac{1-P_+}{P_+}) \quad (75)$$

For second order reaction.

$$\frac{1}{C} = \frac{1}{C_0} + kt;\ kt = \ln(\frac{1-P_+}{P_+}) \quad (76)$$

For decreasing reactions is similar.

## 4. Meaning and application of Corollary 5, or stability and convergence.

The results obtained for the different diffusion differential equations for the temperature of heat equation and diffusion concentration are valid when the time tends to infinity t→∞.

In addition to the results of Corollary 5, it must be shown that the probability values converge when the time tends to infinity, so:

$$\lim_{t \to \infty} P_- = \lim_{t \to \infty} \frac{1}{1 + e^{-kt}} = 1 \quad (77)$$

$$\lim_{t \to \infty} P_+ = \lim_{t \to \infty} \frac{1}{1 + e^{+kt}} = 0 \quad (78)$$

In this way we can verify that our solutions and balances are stable and do not have asymptotes or points of divergence.

All demonstrations given in this paper hold for positive times, t≥0.

## 5. Meaning Table 1.

In physics, the fundamental variable of the interaction of radiation with matter is the effective section σ, which have functional correspondence with the frequency parameter collisions A of the Arrhenius equation. In our experiments, evidence of the variability of A are obtained in the Arrhenius equation, A ≠ constant. This variability of A = Ai supports our affirmation and could be modeled as follows:

$$k = A_i e^{\frac{E_a}{RT}} = \sigma A e^{-\frac{E_a}{RT}} \tag{79}$$

## 6. Meaning Table 2.

Guess Nakamura[9], pag 184, on the realization of the tunnel effect at high temperatures is experimentally verified

**First Example.**

Preliminary tests were performed to establish a working range where there is greater influence of the concentration of methanol in aqueous solution on hydrogen production; the working range for the preliminary tests is defined taking into account as endpoints: pure methanol and pure water, while the intervals were established by researchers. The molar concentrations of hydrogen produced for each concentration of reagent fed were recorded, in addition the curve area of the electrical signals chromatographs obtained for each concentration was recorded, as follows:

| $[CH_3OH]H$, %V/V | $H_2$, % mol | Curve Area, µV*s |
|---|---|---|
| 0 | 29,498 | 37122,7 |
| 10 | 35,697 | 44922,3 |
| 20 | 42,161 | 53055,2 |
| 30 | 25,265 | 31797 |
| 40 | 45,827 | 57667,2 |
| 50 | 25,57 | 32179.9 |
| 60 | 28,821 | 36270,5 |
| 70 | 27,855 | 35055.8 |
| 80 | 26,405 | 33230,9 |
| 90 | 26,811 | 33741,7 |
| 99 | 29,718 | 37399,8 |

**Table 5.** Preliminary tests hydrogen production

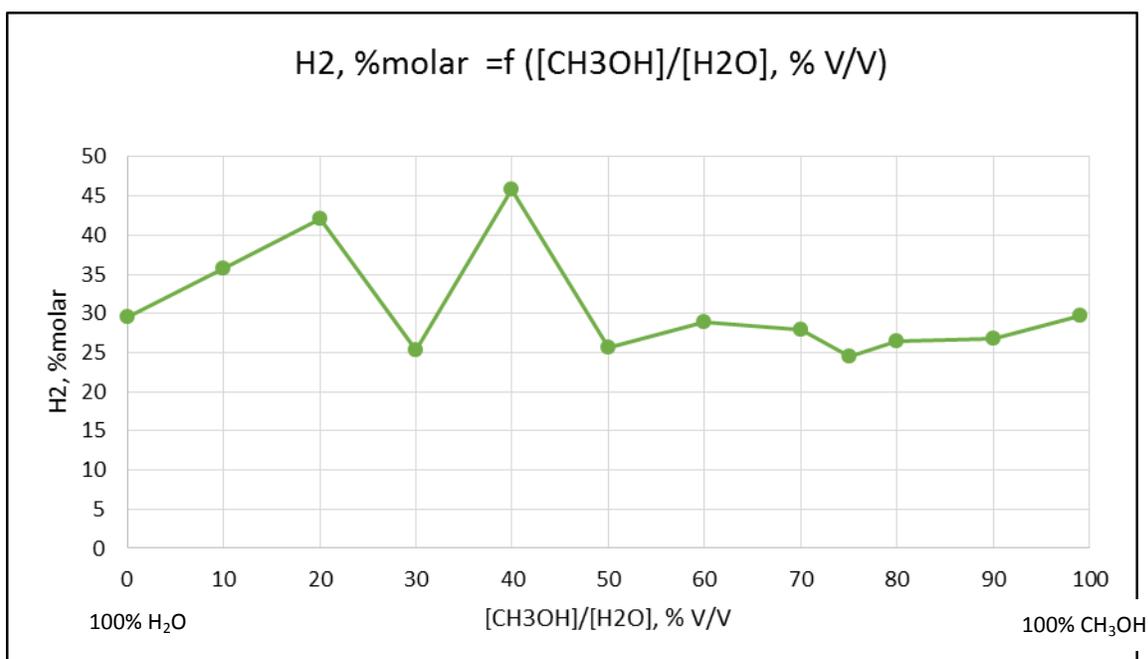

**Figure 6**. Preliminary tests of hydrogen obtained.

As shown in Figure 6, two peaks higher hydrogen production were obtained at 40 and 20% V/V methanol in aqueous solution; for this reason it is taken as a reference to these two peaks to establish a working range from 15 to 45% V/V of methanol in aqueous solution with a variation of 5% for the purpose of expand the points of analysis and obtain a reproducibility process.

The molar percent of hydrogen obtained from 50 to 99.9% V/V methanol in aqueous solution, does not produce a significant variation and these results take a linear trend; also in this range there is no value higher percentage of hydrogen, for this reason are not taken as a point of analysis.

**Working Range.**

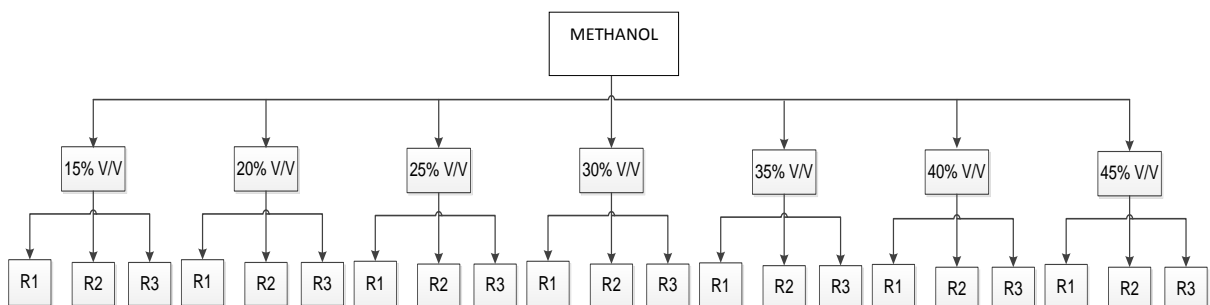

**Figure 7.** Experimental design for the range (15-45% V/V).

For each concentration of methanol in aqueous solution, three replicates of the experiments was carried out, with which ANOVA statistical analysis was performed to determine the influence of the independent variable on the dependent variable.

**Reproducibility Process.**

| [$CH_3OH$], %V/V | Replica 1 | | Replica 2 | | Replica 3 | |
|---|---|---|---|---|---|---|
| | $H_2$, % mol | Curve Area, µV*s | $H_2$, % mol | Curve Area, µV*s | $H_2$, % mol | Curve Area, µV*s |
| 0 | 29,734 | 37418,9 | 31,301 | 39391,4 | 29,824 | 37532,4 |
| 15 | 19,692 | 24785,2 | 22,826 | 28728,4 | 18,994 | 23907,3 |
| 20 | 42,703 | 53736,7 | 44,345 | 55803 | 46,453 | 58455,1 |
| 25 | 36,074 | 45395,9 | 30,563 | 38463,1 | 37,395 | 47058 |
| 30 | 21,298 | 26805,7 | 22,905 | 28827,1 | 25,978 | 32694 |
| 35 | 14,393 | 18117,6 | 13,461 | 16945,8 | 12,613 | 15879 |
| 40 | 45,802 | 57653,8 | 41,971 | 52815,2 | 45,816 | 57653,4 |
| 45 | 17,488 | 22011,5 | 19,41 | 24429,9 | 17,859 | 22478,2 |

**Table 6.** Reproducibility of the experiment in hydrogen production

We can see a chromatographic example in the next figure 8

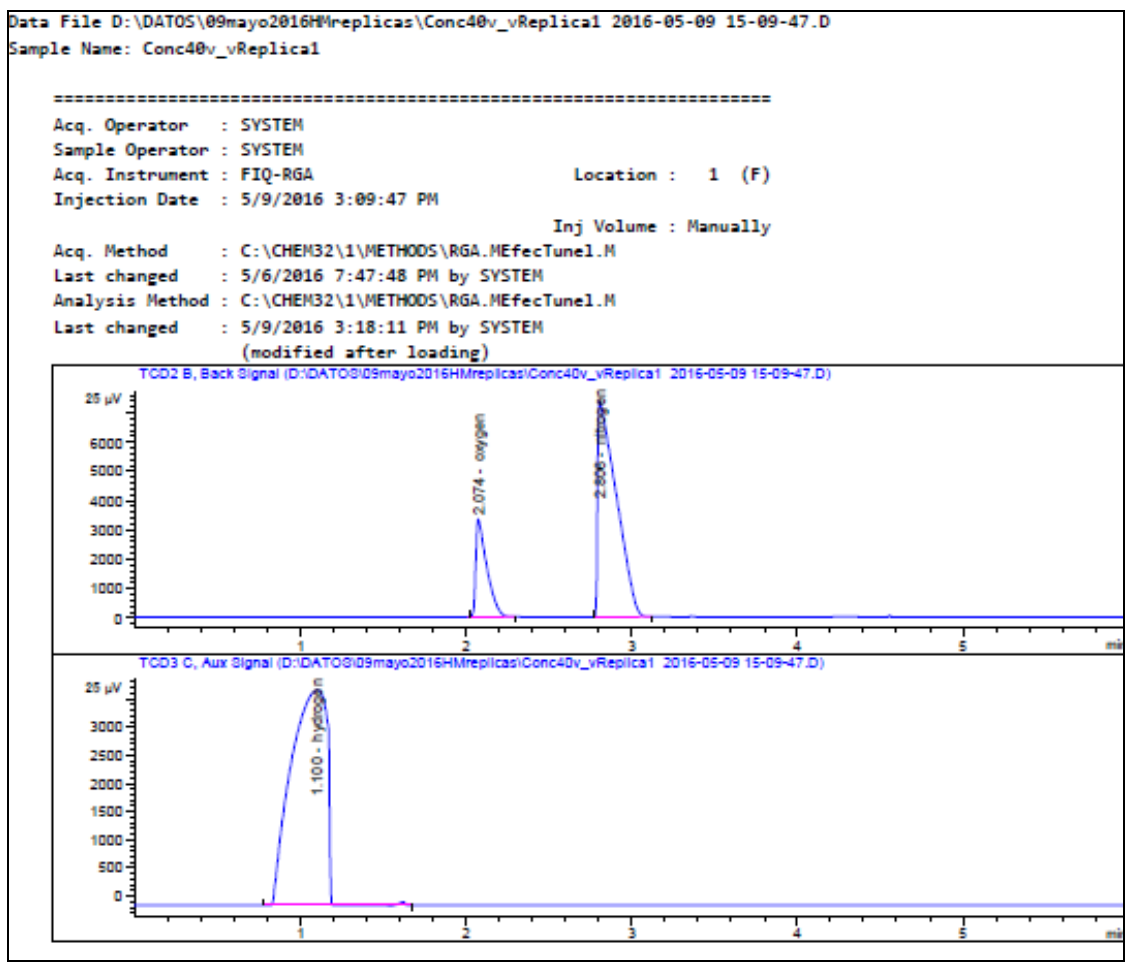

**Figure 8.** Chromatography at 40% v / v of methanol

In this graph the chromatographic signal is observed reporting equipment AGILENT TECHNOLOGIES 7890B (RGA), properly stored in the system

Important to mention that replicas were made for the point of 0% V/V of methanol in aqueous solution to confirm that the molar percentage of hydrogen obtained is real, You can see in figure()

**ANOVA Analysis.**

With replicas obtained a statistical treatment of results was performed using the method of analysis of variance (ANOVA), with a range of acceptance of 95%.
For analysis, the computer tool Microsoft Excel 2013 was used.
Hypotheses were established:

- Null hypothesis: The sample means of the molar percentage of hydrogen obtained are equal to each reactant concentration methanol.

- Alternate Hypothesis: At least one of the sample means of the molar percentage of hydrogen obtained differs from other.

To accept or reject the null or Alterna hypothesis, was analyzed as follows:

| Groups | Count | Sum | Average | Variance |
|---|---|---|---|---|
| 15%V/V | 3 | 61,514 | 20,505 | 4,165 |
| 20%V/V | 3 | 133,503 | 44,501 | 3,534 |
| 25%V/V | 3 | 104,033 | 34,678 | 13,129 |
| 30%V/V | 3 | 70,182 | 23,394 | 5,655 |
| 35%V/V | 3 | 40,469 | 13,490 | 0,792 |
| 40%V/V | 3 | 133,604 | 44,535 | 4,929 |
| 45%V/V | 3 | 54,757 | 18,252 | 1,040 |

**Table 7.** Variance analysis

| Origin of variations | Sum of squares | Degrees of freedom | Average square | F | Probability | Critical value for F |
|---|---|---|---|---|---|---|
| Between groups | 2914,883 | 6,000 | 485,814 | 102,290 | 0,000 | 2,848 |
| Within groups | 66,490 | 14,000 | 4,749 | | | |
| Total | 2981,372 | 20,000 | | | | |

**Table 8.** Analysis of variance - F Fisher

According to these results, it has to F>Fc therefore reject the Null Hypothesis and Alternative Hypothesis is accepted.

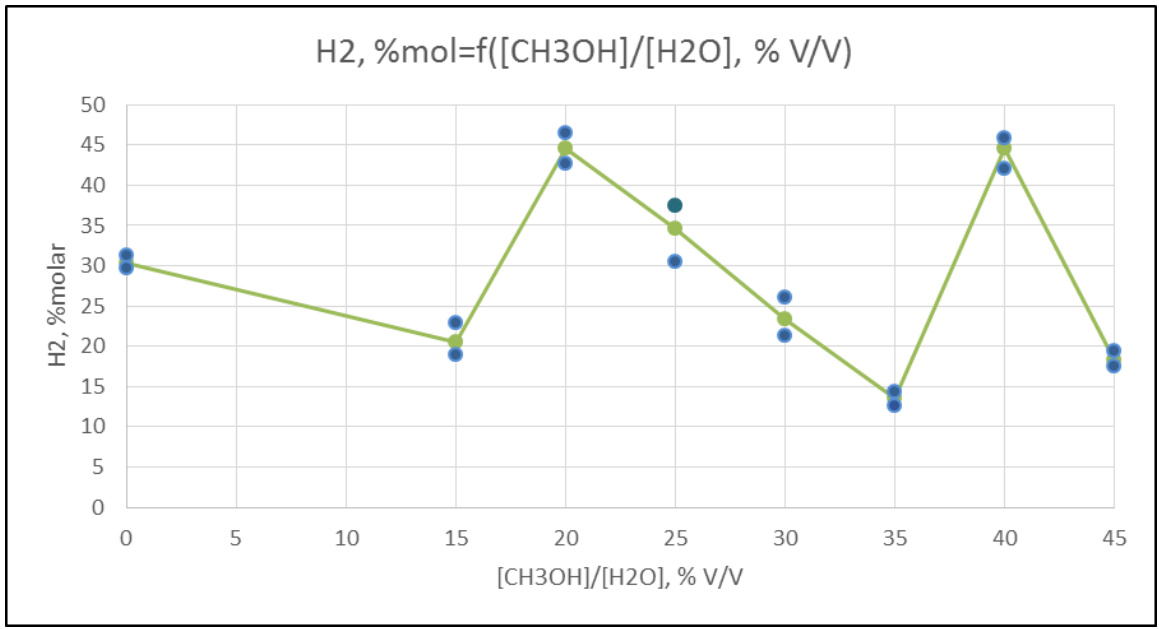

**Figure 9.** Experimental results for obtaining hydrogen.

On this plot we see the molar percentage of hydrogen obtained depending on the methanol concentration using the results of the sample means obtained in ANOVA analysis. It is evident that maximum and minimum points in each replica made, are not significantly away from the sample means.

**Second Example.**

Both the temperature and the catalyst / asphalt directly influence the recovery rate of light compounds.

The temperature presented a quadratic behavior, described as follows, as the temperature increases the recovery rate also increases, reaching a peak, located at 420 ° C, then this vertex the recovery rate decreases.

This is because the increase in temperature produces a higher coke formation product of parallel reactions, quickly dirtying catalyst active sites.

With respect to the mass ratio of catalyst / asphalt, this likewise presented a quadratic behavior, resulting in maximum 0.05 percent recovery.

# FCC CATALYST REGENERATION

## BET surface area

| Catalyst FCC | N° | Area m2/cell | Area m2/g | Area Average m2 / g |
|---|---|---|---|---|
| Spent | 1 | 18,82 | 131,241 | 131,19 |
|  | 2 | 19,96 | 131,143 |  |
|  | 3 | 18,85 | 131,176 |  |
| Regenerate | 1 | 21,97 | 149,659 | 149,44 |
|  | 2 | 22,01 | 149,830 |  |
|  | 3 | 20,24 | 148,824 |  |

**Table 9.** FCC catalyst BET surface area

The regenerated catalyst FCC unit has a higher surface area to the spent catalyst; this is mainly due to coke elimination in the regeneration step.

## Size of particle

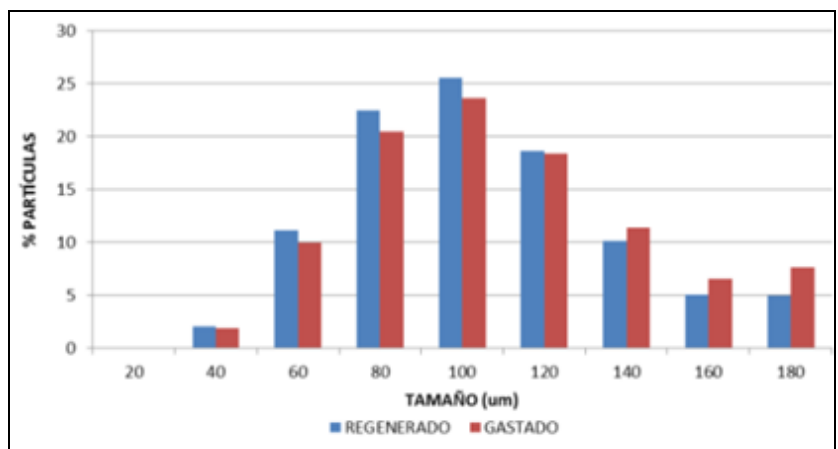

**Figure 10.** Histogram particle size distribution

It is shown in the histogram distributions of particle sizes, that the highest percentage of retention is 80 to 120 μm for the two cases; on the other hand, for particles larger than 120 μm the retention rate of spent catalyst is greater than the percentage retention regenerated catalyst, this is due to decoking and poisonous layers in the regeneration process.

**Fourier Transform Infrared Spectrometry**

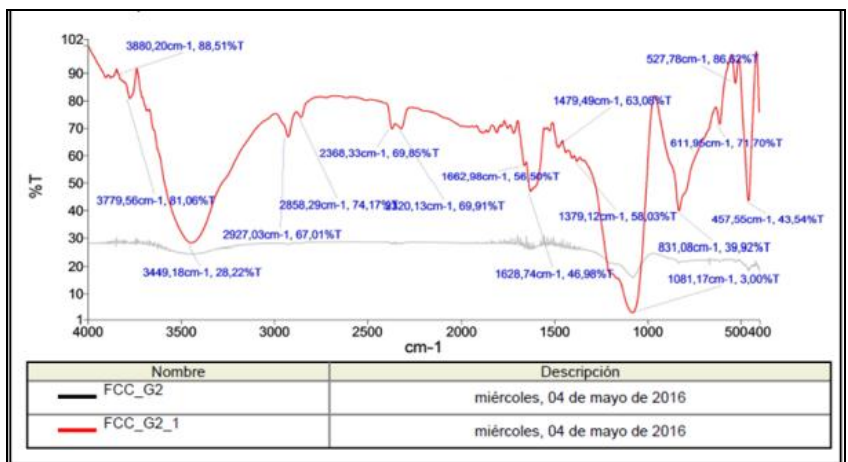

**Figure 11.** Spectrum of spent catalyst

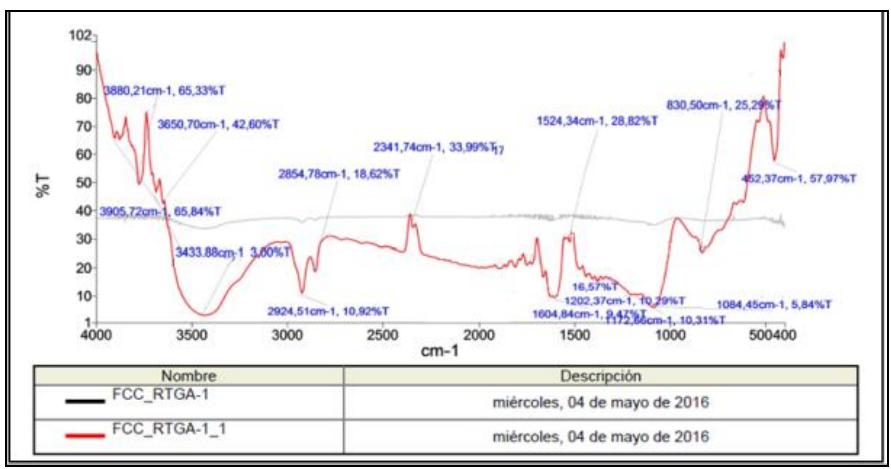

**Figure 12.** Spectrum regenerated catalyst

Comparing the infrared spectra, it is shows that for the two cases there is water presence, oxides of aluminum, silicon.

The bands located at 3400 cm-1 indicate that there was moisture in the spent and regenerated catalyst.

For both cases a peak at 1800cm-1, which corresponds to V5 +, with a lower percentage of transmittance of 7%, this percentage indicates that the vanadium concentration in the FCC catalyst is high, being the main cause of it had catalyst poisoning.

**Percentage of coke combusted**

| N° | Spent Catalyst | Regenerated Catalyst | % Coke Combusted |
|---|---|---|---|
| 1 | 2,5072 | 2,4832 | 0,957 |
| 2 | 10,9422 | 10,8342 | 0,987 |
| 3 | 8,7479 | 8,6625 | 0,976 |
| 4 | 2,4927 | 2,4697 | 0,923 |
| 5 | 7,1025 | 7,0312 | 1,004 |
|  |  | Average | 0,969 |

**Table 10.** Percentage of coke combusted

Depending on the feed quality, the catalyst enters the regenerator containing 0,5 to 1,5% coke. For our case an average value of 0,97 was obtained which indicates that the regeneration temperature was correct. It should be emphasized that the spent catalyst contains impurities such as sodium, vanadium, nickel, zinc oxides, among others which also react in the regeneration process however, concentrations are very small, and so for practical purposes it was assumed that all the mass loss is due to the combustion of coke.

**CATALYTIC PROCESS**

**Comparison reaction temperatures.**

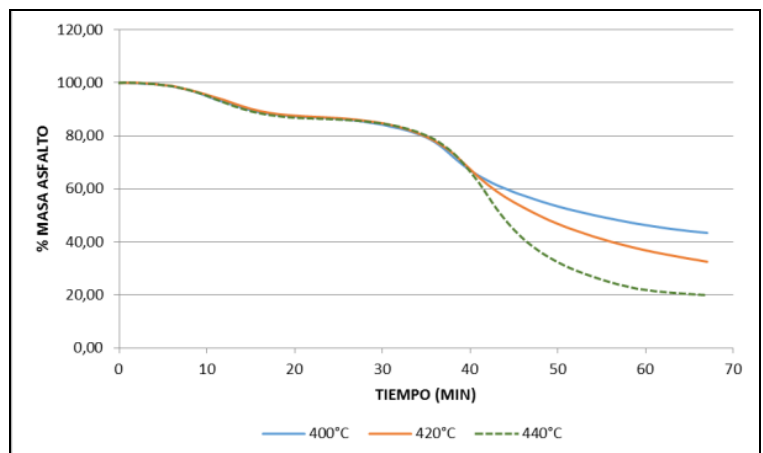

**Figure 13.** Curve thermogravimetric asphalt without catalyst reaction temperatures

In all three cases the asphalt undergoes a thermal cracking followed by evaporation of the residue, this in terms of quality is not favorable since higher percentage of coke plus low octane were generated. Most gas generation rate was obtained at 440 °C temperature.

**Thermogravimetric curves for each catalyst/Asphalt ratio and each reaction temperature.**

Each curve corresponds to the average of three trials.

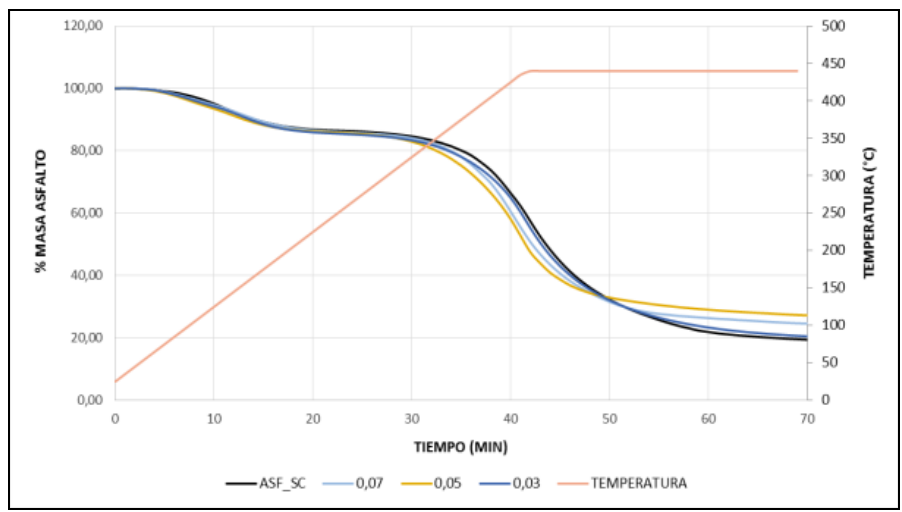

**Figure 14.** Thermogravimetric curve at 400 ° C

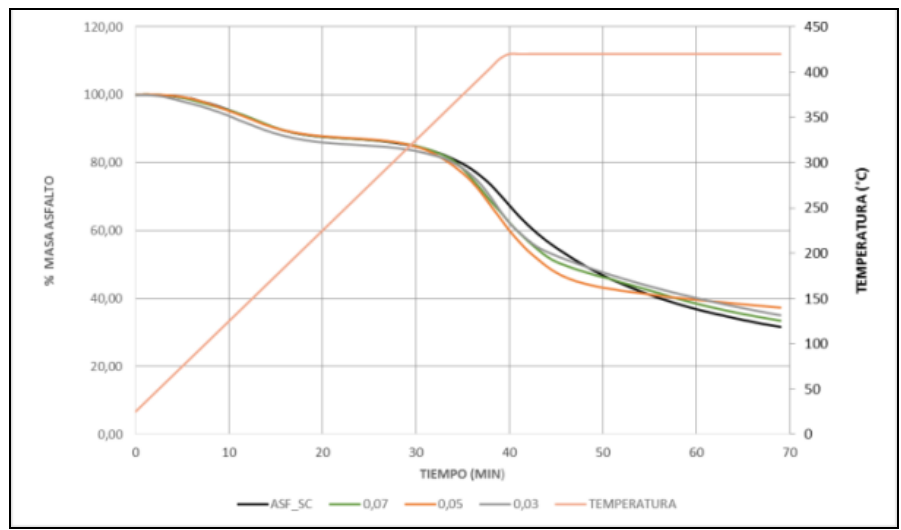

**Figure 15.** Thermogravimetric curve at 420 ° C

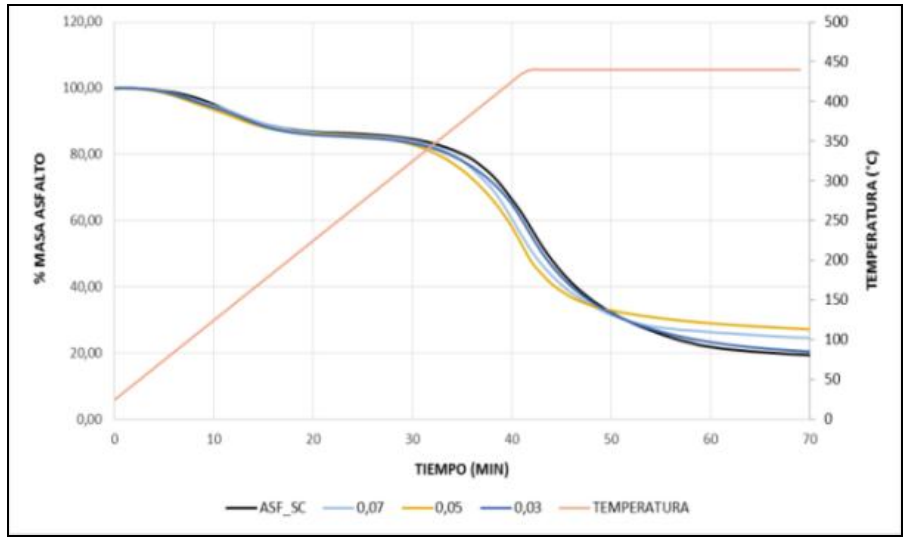

**Figure 16.** Thermogravimetric curve at 440 ° C

By comparing the asphalt curve without catalyst with each of the relations, figures 14, 15, and 16 three zones are viewed, the first zone corresponds to moisture loss and evaporation of small amounts of light compounds, the second corresponds to thermal cracking and catalytic cracking respectively and the third zone corresponds to the residue's evaporation.

| Temperature °C | Relations (Mass Catalyst / Mass Asphalt) | | |
|---|---|---|---|
| | 0,03 | 0,05 | 0,07 |
| 400,00 | 3,80 | 7,32 | 1,86 |
| | 3,82 | 6,93 | 1,77 |
| | 3,65 | 7,70 | 1,67 |
| 420,00 | 4,29 | 10,38 | 5,01 |
| | 5,24 | 11,10 | 4,96 |
| | 4,27 | 10,66 | 6,89 |
| 440,00 | 1,31 | 5,59 | 2,43 |
| | 1,28 | 5,33 | 3,10 |
| | 1,52 | 6,56 | 2,67 |

**Table 11.** Percent Recovery

**Analysis of the gaseous products**

|  | Without catalyst | Regenerated catalyst |
|---|---|---|
| Temperature °C | 420 | 420 |
| Ratio *m/m* catalyst/asphalt | 0,00 | 0,05 |
| Composition gaseous products, %mol | | |
| Hydrogen | 0,020 | 0,094 |
| Methane | 0,797 | 1,311 |
| Ethane | 0,930 | 0,950 |
| Ethylene | 0,485 | 0,749 |
| Propane | 1,404 | 1,766 |
| Propylene | 0,924 | 2,664 |
| iso-Butane | 0,215 | 0,544 |
| n-Butane | 1,578 | 1,906 |
| Propadiene | 0,000 | 0,000 |
| Acetylene | 0,000 | 0,010 |
| trans-2-Butene | 0,552 | 1,940 |
| 1-Butene | 0,123 | 0,100 |
| iso-Butylene | 0,347 | 1,183 |
| cis-2-Butene | 0,342 | 1,031 |
| iso-Pentane | 2,487 | 3,295 |
| n-Pentane | 3,044 | 3,122 |
| 1,3-Butadiene | 0,000 | 0,007 |
| trans-2-Pentene | 1,456 | 2,796 |
| 2-Methyl-2-butene | 0,592 | 0,729 |
| 1-Pentene | 1,604 | 1,841 |
| cis-2-Pentene | 1,936 | 2,053 |
| n-Hexane | 2,528 | 2,878 |
| C6+ | 78,636 | 69,031 |
| **TOTAL** | **100,000** | **100,000** |

**Table 12.** Composition of the gaseous products

The analysis of gases produced in catalytic cracking was performed in a chromatograph for analysis of refinery gas RGA (Agilent Technologies 7890A), in Table 8 the obtained results are detailed, which shows that has a high percentage of major compounds to six carbon atoms, 78.63% for asphalt and 69,03% for asphalt-catalyst mixture, this is because the asphalt is a complex mixture of compounds of high molecular weight which are more difficult to crack. In addition, that it has great amounts of heavy metals which poison active sites of catalyst FCC, rapidly decreasing its activity.

In the catalytic cracking occurred several reactions, which were from the process, the results in Table 12 show an increase in the percentage of hydrogen, methane, propane, n-butane, propylene, i-butane, trans-2-butene, trans 2-pentene, this is mainly due to isomerization reactions and disintegration.

**REMARKS**

Based on the results of BET surface area, particle size, percentage of coke burned and mainly the percent recovery of lighter compounds, we can concluded that regenerated catalyst possesses catalytic activity. Experimentally, it was shown that the temperature of 420 °C and the catalyst/asphalt ratio of 0,05 provided the best catalytic performance in terms of production of cracked gases. Phenomenon properly explained by the Chaos Theory. Minimum catalyst's concentrations produce maximum catalytic cracking. Verifying once again that there is controlled chaos which helps in catalyzing the reaction; thus changing the composition of the asphalt prior to the nanotechnological treatment.

**CONCLUTION**

**FIRST EXAMPLE**

**We obtain two local optima of hydrogen production at the concentrations of methanol in aqueous solution of 20 and 40% V / V.**

- At concentrations of methanol in water %V/V between [10%, 50%] we obtained two local maxima and one minimum in hydrogen production.

- The maximum value of hydrogen production implies existence of cooperative phenomena and correlation, above baseline levels.

- The minimum value in the production of hydrogen implies existence of phenomena of noncooperation or non-correlation.

This new method in hydrogen production will provide a cost-effective production since current methods are based on the classical chemical, which leads to operate with high temperatures and pressures that implies huge cost; However, thanks to quantum chemistry we can choose for another and more viable process, which operates with interstellar temperatures (temperatures extremely low 82±1 K). This research could benefit oil and chemical industry with implementation of new operating conditions more viable and cost-effective hydrogen production methods

**SECOND EXAMPLE**

- The results of surface area, particle size distribution, infrared spectroscopy and percentage of coke burned, corroborates that produced a correct regeneration of spent FCC catalyst unit.

- Catalytically cracking was achieved asphalt, obtaining a maximum recovery rate of 10% lightweight composite at a temperature of 420 ° C and a mass ratio of catalyst / asphalt 0.05 [$g_{cat}/g_{asf}$ ].

- The change in the composition of the gaseous products confirms that the asphalt suffered a catalytic cracking, since an increase in the percentage of hydrogen, propane, i-butane, n-butane, propylene, trans-2-butene, was obtained trans- 2-pentene.

- These results indicate that a tunnel effect was given to high temperatures.

- Phenomenon properly explained by the Chaos Theory, minimum catalyst's concentrations produce maximum catalytic cracking. Verifying once again that there is controlled chaos which helps in catalyzing the reaction; thus changing the composition of the asphalt prior to the nanotechnological treatment.

# REFERENCES

1. Kolmogorov A, Petrovsky I and. Piscounov N. Study of the diffusion equation with growth of the quantity of matter and its application to a biological problem, Bull. State Univ. Mos, (trans. by F. Oliveira- Pinto and B. W). *Conolly, Applicable mathematics of non-physical phenomena, Ellis Horwood*, 1937, *1982*, 169-184.

2. Bokshtein, B, Mendelev S, Srolovitz M.I. D.J., *Thermodynamics and Kinetic in Material Science, Oxford University Press*, 2005, pp. 167-171

3. Nakamura H, Mil'nikov G. *Quantum Mechanical Tunneling in Chemical Physics*. CRC Press Taylor & Francis Group, Boca Raton, Florida, 2013, pp. 17-19

4. Bell R. P, *The Tunnel Effect in Chemistry*, Springer U.S, 2013.pp. 1-11 ; 32-62

5. Lim, T. Long range relationship between Morse and Lennard–Jones potential energy functions. *Molecular Physics*, 2007, *105(8),* pp.1013-1018.

6. Tannoudji C et al. *Quantum Mechanics*, John Wiley & Sons, New York, 1977, pp. 509-600.

7. Griffiths, David J. *Introduction to Quantum Mechanics (2nd ed.). Prentice Hall.* 0-13-805326-X. 2014, pp. 133-200

8. Palmer, T. A Local Deterministic Model of Quantum Spin Measurement. *Proceedings of the Royal Society A: Mathematical, Physical and Engineering Sciences*, 1995 *451(1943),* pp.585-608.

9. Weiss, U. *Quantum dissipative systems*. Singapore: World scientific, 1999 Vol. 13, pp. 47-53.

10. Smith, I. The temperature-dependence of elementary reaction rates: beyond Arrhenius. *Chem. Soc. Rev*., 2008, *37(4),* pp.812-826.

11. Chacon, A. and Vladimirsky, A. A Parallel Two-Scale Method for Eikonal Equations. *SIAM J. Sci. Comput*., 2015, *37(1),* pp.A156-A180.

12. Oliveira, R. I. Concentration of the adjacency matrix and of the Laplacian in random graphs with independent edges. *arXiv preprint arXiv:0911.0600,* 2009,pp 1-46

13. Mahan, G. The tunneling of heat. *Appl. Phys. Lett.,* 2011. *98(13),* p.132106.

14. Yong-Wang L. Clean Diesel Production from Coal Based Syngas via Fischer-Tropsch Synthesis: Technology Status and Demands in China. *Chinese Academy*, 2014, pp. 1-13.

15. Bell, R. and Le Roy, R. The Tunnel Effect in Chemistry. *Phys. Today*, 1982, *35(5),* p.85.

16. Compton, R.G. *Electron Tunneling in Chemistry*, Elsevier: New York, 1989, Vol. 30, pp. 121-125

17. Li, Z., Wang, Y., Liu, J., Cheng, G., Li, Y. and Zhou, C. Photocatalytic hydrogen production from aqueous methanol solutions under visible light over Na(BixTa1−x)O3 solid-solution. *International Journal of Hydrogen Energy*, 2009, *34(1),* pp.147-152.

18. Ley, D., Gerbig, D. and Schreiner, P. Tunneling control of chemical reactions: C–H insertion versus H-tunneling in tert-butylhydroxycarbene. *Chem. Sci.*, 2013, *4(2),* pp.677-684.

19. Shannon, R., Blitz, M., Goddard, A. and Heard, D. Accelerated chemistry in the reaction between the hydroxyl radical and methanol at interstellar temperatures facilitated by tunnelling. *Nature Chemistry*, 2013, *5(9),* pp.745-749.

20. Silva, V., Aquilanti, V, de Oliveira, H. and Mundim, K.. Uniform description of non-Arrhenius temperature dependence of reaction rates, and a heuristic criterion for quantum tunneling vs classical non-extensive distribution. *Chemical Physics Letters*, 2013, *590*, pp.201-207.

21. Carrera, D. Quantum Tunnelling in Chemical Reactions. *MacMillan Group Meeting.* 2007.Online at: https://www.princeton.edu/chemistry/macmillan/group-meetings/DEC_tunneling.pdf

22. Compton R.G. (Editor). *Electron Tunneling in Chemistry: Chemical Reactions over Large Distances Comprehensive Chemical Kinetics*. Elsevier, Amsterdam, 1989, Vol. 30, pp.45-63

23. Bičáková, O. and Straka, P. Production of hydrogen from renewable resources and its effectiveness. *International Journal of Hydrogen Energy*, 2012, *37(16),* pp.11563-11578.

24. Amphlett, J., Evans, M., Mann, R. and Weir, R. Hydrogen production by the catalytic steam reforming of methanol: Part 2: Kinetics of methanol decomposition using girdler G66B catalyst. *Can. J. Chem. Eng.*, 1985, *63(4),* pp.605-611.

25. Kreuer, K. On the development of proton conducting polymer membranes for hydrogen and methanol fuel cells. *Journal of Membrane Science*, 2001, *185(1),* pp.29-39.

26. Tedsree, K., Li, T., Jones, S., Chan, C., Yu, K., Bagot, P., Marquis, E., Smith, G. and Tsang, S. Hydrogen production from formic acid decomposition at room temperature using a Ag–Pd core–shell nanocatalyst. *Nature Nanotech*, 2011, *6(5),* pp.302-307.

27. López Ortiz, A. and Collins Martínez, V. Preface to the special issue on "XIV International Congress of the Mexican Hydrogen Society, 30 September–4 October 2014, Cancun, Mexico". *International Journal of Hydrogen Energy*, 2015, *40(48),* pp.17163-17164.

28. Félix Flores, Simulación del proceso FCC: Caracterización de las Corrientes de Alimentación y Productos del Riser. Ph.D en Ciencias en Ingeniería Química Instituto Tecnológico de Celaya, México, 2008.

29. Sadeghbeigi R. *Fluid Catalytic Cracking Handbook*, Elsevier Inc, Houston. TX, 2012, pp. 87-113.

30. Harding, R., Peters, A., & Nee, J. New developments in FCC catalyst technology, *Applied Catalysis*, 2001, 221, 389.

31. Fogler, H. S., *Elementos de Ingeniería de las Reacciones Químicas*, Pearson Educación, México, 2008, pp. 581-665

32. Hernández Barajas, J. Simulación Dinámica del Proceso FCC: Una Propuesta Cinética Basada en Distribuciones de Probabilidad, Ph.D, Instituto Tecnológico de Celaya, México, 2003.

33. Fang, C., Yu, R., Liu, S. and Li, Y. Nanomaterials Applied in Asphalt Modification: A Review. *Journal of Materials Science & Technology*, 2013. *29(7),* pp.589-594.

34. Gamarra, A., & Ossa, A. Efecto de la foto-degradación en la microestructura y composición química del asfalto, *Revista Colombiana de Materiales*, N. 5, 7-12.

35. Fang, C., Wu, C., Hu, J., Yu, R., Zhang, Z., Nie, L., Zhou, S. and Mi, X. Pavement properties of asphalt modified with packaging-waste polyethylene. *J Vinyl Addit Technol*, 2014, *20(1)*, pp.31-35.

36. Fesharaki, M. J., Ghashghaee, M., & Karimzadeh, R., *Journal of Analytical and Applied Pyrolysis*, 102, 97-102.

37. Moreno Mayorga, J. C., Ancheyta Juárez, J., & Gamero Melo., *Journal Of the Méxican Chemical Society*, 41.

38. Souza, J., Vargas, J., Ordoñez, J., Martignoni, W., & Von Meien, O., *International Journal of Heat and Mass Transfer*, 54.

39. Kuznicki, S., McCaffrey, W., Bian, J., Wangen, E., Koenig, A., & C.H. Lin, C., *Microporous and Mesoporous Materials*, 105, 268–272.

40. Abu S.M. Junaid, Moshfiqur Rahman, Haiyan Yin, William C. McCaffrey, Steven M. Kuznicki, Natural zeolites for oilsands bitumen cracking: Structure and acidity, *Microporous and Mesoporous Materials*, 144, 148-157

41. Goldstein H, Poole Ch and Safko J. *Classical Mechanics*, Addison Wesley, San Francisco, 2001, pp. 483-525

42. Atkins P and de Paula J, *Physical Chemistry*,W.H. Freeman and Company, New York, 2006, pp.870-880